\documentclass[
aps, pra,
amsmath,amssymb,
11pt,
tightenlines,
twoside,
twocolumn,
nofootinbib,
superscriptaddress,
showkeys,
]{revtex4-2}
\usepackage[pdftex]{graphicx}
\usepackage{booktabs}
\usepackage{longtable}
\usepackage{wasysym}
\def\saoname{Special Astrophysical Observatory,  Russian Academy of Sciences,
              Nizhnii Arkhyz, 369167 Russia}

\def\squareforqed{\hbox{\rlap{$\sqcap$}$\sqcup$}}

\def\sq{\ifmmode\squareforqed\else{\unskip\nobreak\hfil
\penalty50\hskip1em\null\nobreak\hfil\squareforqed
\parfillskip=0pt\finalhyphendemerits=0\endgraf}\fi}

\def\utw{\smash{\rlap{\lower5pt\hbox{$\sim$}}}}

\def\udtw{\smash{\rlap{\lower6pt\hbox{$\approx$}}}}

\def\diameter{{\ifmmode\mathchoice
{\ooalign{\hfil\hbox{$\displaystyle/$}\hfil\crcr
{\hbox{$\displaystyle\mathchar"20D$}}}}
{\ooalign{\hfil\hbox{$\textstyle/$}\hfil\crcr
{\hbox{$\textstyle\mathchar"20D$}}}}
{\ooalign{\hfil\hbox{$\scriptstyle/$}\hfil\crcr
{\hbox{$\scriptstyle\mathchar"20D$}}}}
{\ooalign{\hfil\hbox{$\scriptscriptstyle/$}\hfil\crcr
{\hbox{$\scriptscriptstyle\mathchar"20D$}}}}
\else{\ooalign{\hfil/\hfil\crcr\mathhexbox20D}}%
\fi}}

\def\No{N\textsuperscript{\underline{o}}}

\begin{document}
\bibliographystyle{unsrtnat}

\keywords{radiocontinuum: stars; stars: flares}

\title{Search for activity in the centimeter band of Solar-type dwarfs with
 Earth-like planets on the RATAN-600 radio telescope.
}
\author{\firstname{G.~M.}~\surname{Beskin}}
 \email{beskin@sao.ru}
 \affiliation{\saoname}

\author{\firstname{V.~N.}~\surname{Chernenkov}}
 \email{vch@sao.ru}
 \affiliation{\saoname}

\author{\firstname{N.~N.}~\surname{Bursov}}
 \email{nnb@sao.ru}
 \affiliation{\saoname}

\author{\firstname{A.~A.}~\surname{Shlyapnikov}}
 \email{aas-crao@mail.ru}
 \affiliation{Crimean astrophysical observatory,  p. Nauchny, Bakhchisaray,
  298409 Crimea}
\author{\firstname{A.~D.}~\surname{Panov}}
 \email{panov@dec1.sinp.msu.ru}
 \affiliation{M.V.Lomonosov Moscow State University,
 Skobeltsyn Institute of Nuclear Physics,
 119991 Russia}

\begin{abstract}
The paper presents the results of multi-wave radio observations of sixteen
red dwarfs with Earth-like planets in the habitable zones on RATAN-600.
In the passage mode, radiometers of four frequency ranges of
22, 11, 4.7 and 2.3 GHz (1.38, 2.68, 6.38, 13.3 cm) were used with a
sampling step of quasi-simultaneous
registration of about $\sim 18$ Hz and an exposure duration (the time of passage
of the instrument's directivity pattern) of 1.5 -- 15 s. Observations were
carried out from 3 to 27 times in April-May 2018, while radio emission was
not detected. Upper limits were established for its intensity in flares with
a duration of 0.05 -- 10 s at the level of 80 -- 800 mJy (corresponding to a
luminosity of $10^{23} - 10^{29}$ erg/s) and in a steady state with a luminosity
of $10^{22} - 10^{27}$ erg/s.
During the observations, 11 anomalous flux outlier at the level of 100 mJy
were registered, which apparently have an atmospheric (thunderstorms,
airplanes) origin.
Assuming an artificial origin of the sought radio emission (transmission
from inhabited planets), the upper limits of the power of hypothetical
transmitters were $2\times 10^9 - 10^{15}$ W,  interestingly that the
lower boundary  of this interval close to the power of terrestrial
planetary radars.
\end{abstract}

\maketitle

\section{Introduction.}
Dwarfs of late spectral classes from F to M are usually classified as
solar-type stars based on the similarity of their activity to that of
the Sun \cite{2003PhyU...46...97S,2019PASP..131a6001M,2024arXiv241111898G}.
In our Galaxy, this sample contains, according to various estimates, about
160 billion objects, which is approximately 80 percent of the entire stellar
population.

The variability of such stars, as well as the Sun, is caused by magnetic
activity, which manifests itself in periodic variations in brightness as a
result of rotation \cite{2002AN....323..157S,2023ApJ...954L..50E} and
sporadic flares - a consequence of the reconnection of magnetic field
force lines \cite{2024LRSP...21....1K}. The magnetic fields themselves
are generated by the combined action of differential rotation and
convection \cite{2017LRSP...14....4B}. The effects of activity are
observed in various spectral ranges from X-rays to radio and, exceeding
the solar level by 100 - 1000 times \cite{2024arXiv241111898G}, can
significantly affect the possibility of the emergence and preservation of
life on planets discovered around red dwarfs \cite{2011ASPC..451..285E}.

The Gaia DR3 catalogue \cite{2023A&A...674A...1G} contains about half a
million solar-type stars, with the largest number of them in this sample
belonging to the spectral class M (about 23\%), and the maximum in the
distribution of stellar magnitudes in the V band falls within the interval
$11^m - 12^m$. The flare with the largest amplitude $\Delta U \approx 12^m.5$
was recorded in the red dwarf V374 Peg $(V = 12^m.01, M3.5Ve)$
\cite{2001AstL...27...29B}.

Studies of stellar flares using modeling within the framework of radiation
hydrodynamics and physical diagnostics of spectra, as well as multi-wave
observations and analysis based on them of processes in stellar atmospheres
have yielded important results in recent years \cite{2024LRSP...21....1K}.

An exceptional role here was presented by observations of the Kepler and
TESS space telescopes, designed to search for exoplanets and, as a
consequence, to study the possible connection between flares and the
emergence and development of life on them. Since these programs used the
transit search method (75\% of detections), which is most sensitive when
observing low-mass red dwarfs \cite{2016ASSL..428...89C}, these objects
were studied in large quantities. In particular, the catalog of observation
results with the Kepler telescope contains 51542 Main Sequence stars of
spectral classes $A-M$, in which 16066 flares were detected. Their maximum
number, 5445, was registered in 2222 M-dwarfs, and the energy of all
events exceeded $10^{33}$ erg \cite{2023RAA....23j5010A}.

Most solar-type stars in the lower part of the Main Sequence show slowly
varying radio emission in the frequency band from 50 MHz to 100 GHz with a
flux in the range of $0.1 - 2.5$ mJy, which can persist for years with some
exceptions (see below), and flare activity on times of seconds - minutes
with a flux up to 10 mJy \cite{2003PhyU...46...97S,2019PASP..131a6001M,
2024arXiv241111898G}.
Both phenomena are based on a combination of auroral emission and rapidly
developing exponentially decaying processes of coronal energy release.

To determine the radio emission in the quiescent state, long-term
observation series and individual flux estimates made in different
epochs (in some cases, the observation intervals were years) were
analyzed \cite{2024arXiv241111898G}. In particular, for 44 stars at 12
frequencies in the range of 857~MHz - 375~GHz in the absence of flares, a
flux at a level of 0.05 to 4 mJy was recorded. For two stars, YZ CMi
(at 375 and 273~GHz) and Gl~867~B (at 5~GHz), the flux was 14~mJy for
individual epochs in the quiescent state, while for Gl~867~B it smoothly
varied from 14 to 0.2~mJy over the course of a month. It was noted
\cite{2024arXiv241111898G} that in some cases the recorded flux from stars
is not the radiation of quiescent coronas, but is caused by individual
active events.

Rapidly growing, exponentially decaying flares of varying amplitudes
have been recorded for most F-G-K-M stars. For example, the flare
activity of 17 Sun-like stars is described in detail in
\cite{2024arXiv241111898G}. Observations were performed in a wide range
from 4.9~MHz to 231~GHz. Flares with amplitudes from 0.4~mJy to 1.6~Jy (!)
were recorded, with characteristic times from 1 microsecond to 3 hours.
In the latter case, such a long-lasting, strongly polarized radiation had
no analogues among solar phenomena. The total energy of the bursts sometimes
exceeded similar events on the Sun by four orders of magnitude, and the
recorded flux was $5-6$ times higher than the level of quiet radio emission.

The data on the radio emission of stars are summarized in the catalog
\cite{2024PASA...41...84D}, which includes 839 objects observed 3405
times, 756 of which are included in the SIMBAD\footnote{SIMBAD Astronomical Database - CDS (Strasbourg)} database. In this sample,
206 stars are dwarfs of spectral classes F-M, for which the minimum flux of
0.18~mJy was recorded at a frequency of 1.4~GHz, and the maximum of 19.2~mJy
at a frequency of 887.5~MHz.

As noted above, the activity of host stars of exoplanets (especially
M-dwarfs) can play an exceptional role both in the origin of life and its
preservation on these planets (see, for example, \cite{2011ASPC..451..285E}
and references therein). Moreover, it is radio emission that reflects the
processes of generation of energetic particles in the stellar wind, which
affect the atmospheres and surfaces of exoplanets \cite{2024arXiv241111898G}.

Cross-identification of the catalog \cite{2024PASA...41...84D} and the NASA
exoplanet database (https://science.nasa.gov/exoplanets/) yielded a match
within $2"$ for 9~stars: 2MASS J21252752-8138278, 2MASS J01033563-5515561,
AB~Pic, AF~Lep, AU~Mic, GJ~1151, GJ~896 A, Proxima Cen, PZ~Tel. This is,
so far, the largest number of host stars of exoplanets with detected radio
emission.

When searching for radio emission at frequencies of 4-8~GHz from 77 stars
with 140 exoplanets, only one star, GJ~3323, recorded a flux at the level
of $86 \pm 10$ $\mu$Jy, corresponding to a luminosity of $\sim 10^{22}$ erg/s;
this value is the upper limit for the emission of the remaining
objects \cite{2024PASA...41...84D}.

In addition to studying radio emissions from stars of natural origin,
searches for radio transmissions from extraterrestrial civilizations have
been conducted for over 60~years, both those directed directly to Earth
and those intended for other civilizations \cite{1961PhT....14d..40D,
1980Icar...42..136T}. The SETI program has not yet yielded the expected
results: no technosignatures have been detected in the radio range (or
in optics) \cite{2013ApJ...767...94S,2017ApJ...849..104E,2020AJ....159...86P}.
In recent years, these observations have been conducted in the gigahertz
range, but low frequencies have been used for the most extensive search
for narrow-band radiation of artificial origin. 1631198 stars from the TESS
and Gaia catalogs were synchronously studied in the 110-190~MHz range at
two international LOFAR stations (in Ireland and Sweden)\cite{2013A&A...556A...2V}.
The data obtained made it possible to set the upper limit for the
transmission power at $10^{17}$W \cite{2023AJ....166..193J}.

In 2015-2016, we conducted a series of regular observations under
the SETI program using the RATAN-600 radio telescope. The objects of
the search for radio emission of artificial origin were about 30 sun-like
stars and two globular clusters with high metallicity. Directional signals,
as well as other technosignatures, were not detected - upper limits were
established for the power of their hypothetical transmitters \cite{panov2019}.

The peculiarity of these studies was multiple repeating sets of
observations (monitoring) of the same objects. This approach is conditioned
by a priori ideas about the strategy of sending signals to other c
ivilizations - transmissions should be regularly repeated.

A similar technique is used in the program, the results of which are
presented in \cite{2023arXiv230301791B} and this paper. We repeatedly observed 16 solar-type
stars (M dwarfs and two G dwarfs) with Earth-like planets, whose orbits are
localized in the habitable zones (see https://phl.upr.edu/hwc) in the
transition mode on the RATAN-600 radio telescope. The registration was
carried out quasi-synchronously at 4 waves (1.38, 2.68, 6.38, and 13.33~cm)
with a time resolution of $\sim 50$~ms.

The purpose of the observations was to search for calm and flare radio
emission from stars to analyze the influence of the processes that generate
it on the physical and chemical conditions in the atmospheres and on the
surface of planets, as well as signals from hypothetical Earth-type
civilizations that presumably inhabit these planets.

Sections \ref{obs} -- \ref{concl}  provide the characteristics of
the equipment, observation mode, and data processing methods,
describe the observed objects, and present the observation results.

\section{Observation mode and equipment.}\label{obs}
Star systems of 16 dwarfs with discovered planets in the supposed zone of
existence of life forms were selected for observations.
The list of stars is given in table \ref{Result} of section \ref{disc}.
It was intended to establish upper limits for the flux density for detecting
the radio emission of the listed stellar systems in the series of
observations on the Southern Sector of RATAN-600.

The observations were carried out in the spring months in the interval
from 04/06/2018 to 05/07/2018 mainly at night.
Observation mode - Southern sector with a flat reflector, movable feed -- secondary mirror
\No2 with daily rolling on rails to combine with other observational programs.
A standard set of ``Eridan-2''\cite{berlin2012} and one addition radiometers
were used, the central wavelengths and bandwidths of which are 1.38, 2.68,
6.38, 13.33 cm and 2.5, 0.8, 0.6, 0.4 GHz, respectively.
The recording was made by the acquisition
system~\cite{Tsybulev2011} in an accelerated mode with a sample rate of
about 18 Hz (time resolution $\sim54.8$ms) for each frequency channel.
Each radiometer used
one horn each, the location and horizontal offsets
from the focus of which are presented in Table \ref{horn}.
A carriage with radiometers during the passage of sources through the
``knife'' pattern the antenna is fixed, that is, the possibility signal
accumulation in a single observation is determined only by its width
in each of the four frequency channels.
Since the mode of operation of the receiving paths is modulation, a separate
record was formed for each of the two half-cycles of the modulation.
The resulting single observation file for each radiometer was the vector
sum of these records.
Separate viewing of each of the summarized records allows you to find
and identify a class of very narrow impulse noise in case of detection
anomalies in statistical properties receiving signal.
\begin{table}
\caption{Position of radiometer horns on secondary mirror carriage.}\label{horn}
\begin{tabular}{l|c|r}
\hline
\begin{minipage}[t]{0.3\columnwidth}
Radiometer, wavelendth $\lambda$ cm.\strut\end{minipage} &
\begin{minipage}[t]{0.3\columnwidth}
Horn position\strut\end{minipage} &
\begin{minipage}[t]{0.3\columnwidth}
Horisontal offset, cm\strut\end{minipage} \\
\hline
1.38 & eastern & -1.35\\
2.68 & west &  5.05\\
6.38 & west &  12.9\\
13.33 & west & 34.3\\
\hline
\end{tabular}
\end{table}

\section{Format of source records and software toolkit.}\label{form}
The purpose of processing the records of this observation cycle was to
identify changes in the statistical properties of noise signal during the
passage of the source through the antenna pattern.
Changes were assessed in relation to the same intervals before and after
this passage. Due to the fact that a possible signal is hidden
behind the usual noise recording, it is important to accurately calculate
the start and end times of the analysis.
Measurements of the variation of the coordinates of the position of the
maximum by strong reference sources showed that the contribution of the
receiver horn setting error, the contribution of noise, atmospheric
refraction, and time synchronization does not exceed 0.1 s, which is
commensurate with the sampling period of the recording.
Some biggest error introduces a shift in the catalog value of the right
ascension $Pm\_RA$ of the studied stars due to proper motion, which we
have taken into account.
We used the precomputed TIMEESH or TIMEWSH, respectively for the passage of
the ``east'' or ``west'' working horns that are included in the header
F-file data acquisition system.  Also in analysis intervals were calculated
using the values of the following parameters: LAMBDA -- central wavelength,
AZIMUTH -- observation azimuth, OBSRA -- right ascension of the source
(excluding proper motion), OBSDEC - declination, XBFWHM - predicted
transit time interval width source through the antenna pattern at half power.

The resulting time shift ($ts_\lambda$) of the moment passing through the
maximum of the
directivity diagram for each recording channel is determined by the following
formula:
\begin{equation}
ts_\lambda = \frac{TS-0.018~Pm\_RA}{15~cos(OBSDEC~\pi/180)}
\label{eq:f1}
\end{equation}
In expression (\ref{eq:f1}) the 18th year is entered
observations relative to the coordinates of RA2000 with the conversion of
the dimension of the values of arcseconds into seconds of time.
The TS value is equal to the TIMEESH or TIMEWSH values, up to the sign:
$+$ or $-$, depending on the AZIMUTH and the location of the horn in the
orientation of the three-mirror observing system, see table \ref{horn}.
As a result, we obtain the following formulas for calculating the average
moments of time of the analyzed sections of the record:
\begin{equation}
t0_\lambda = 3600~OBSRA - 3~XBFWHM_\lambda + ts_\lambda
\label{eq:t0}
\end{equation}
\begin{equation}
t1_\lambda = 3600~OBSRA + XBFWHM_\lambda + ts_\lambda
\label{eq:t1}
\end{equation}
\begin{equation}
t2_\lambda = 3600~OBSRA + 3~XBFWHM_\lambda + ts_\lambda
\label{eq:t2}
\end{equation}
Obviously, that the initial and final moments of the analysis
are obtained
by subtracting and adding to the calculation centers the value of the
half-width of the antenna pattern: $XBFWHM_\lambda$. The given data,
together with the original multi-frequency files, are the input for the
FADPS-based~\cite{fadps1993} software package.

\section{Computational procedures.}\label{comp}
\begin{figure}
\centerline{\vbox{\includegraphics[width=8.0cm]{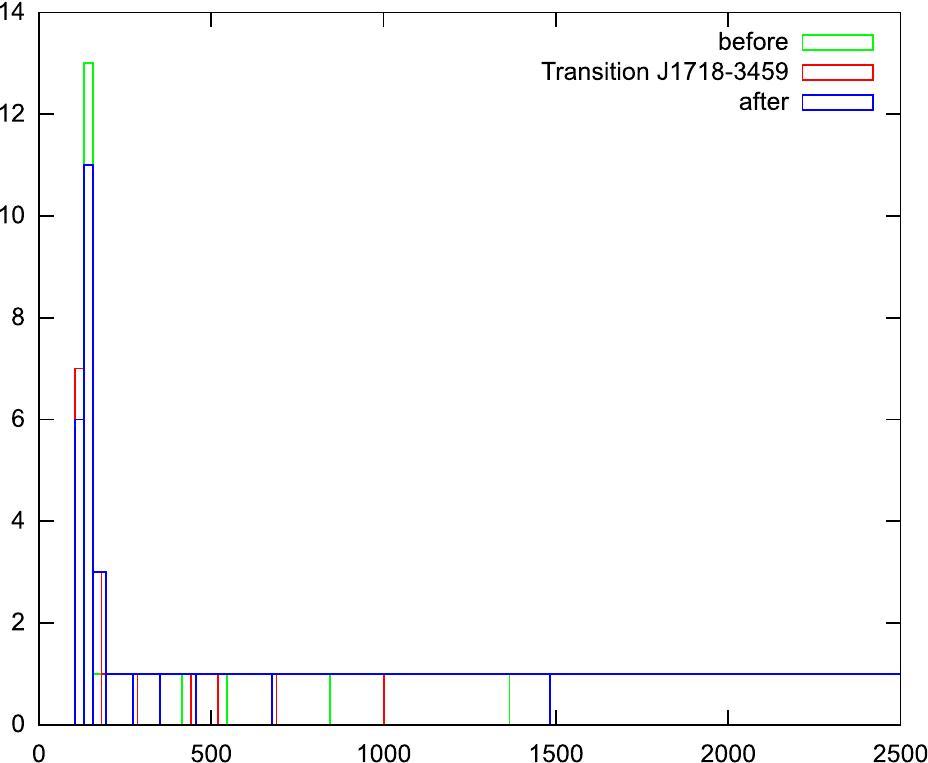}}}
\caption{PDD histogram for a sample of $\sigma$ (mK) record of
GJ667C at 13.33 cm from 26 observations.}\label{histo}
\end{figure}

\begin{figure}
\centerline{\vbox{\includegraphics[width=8.0cm]{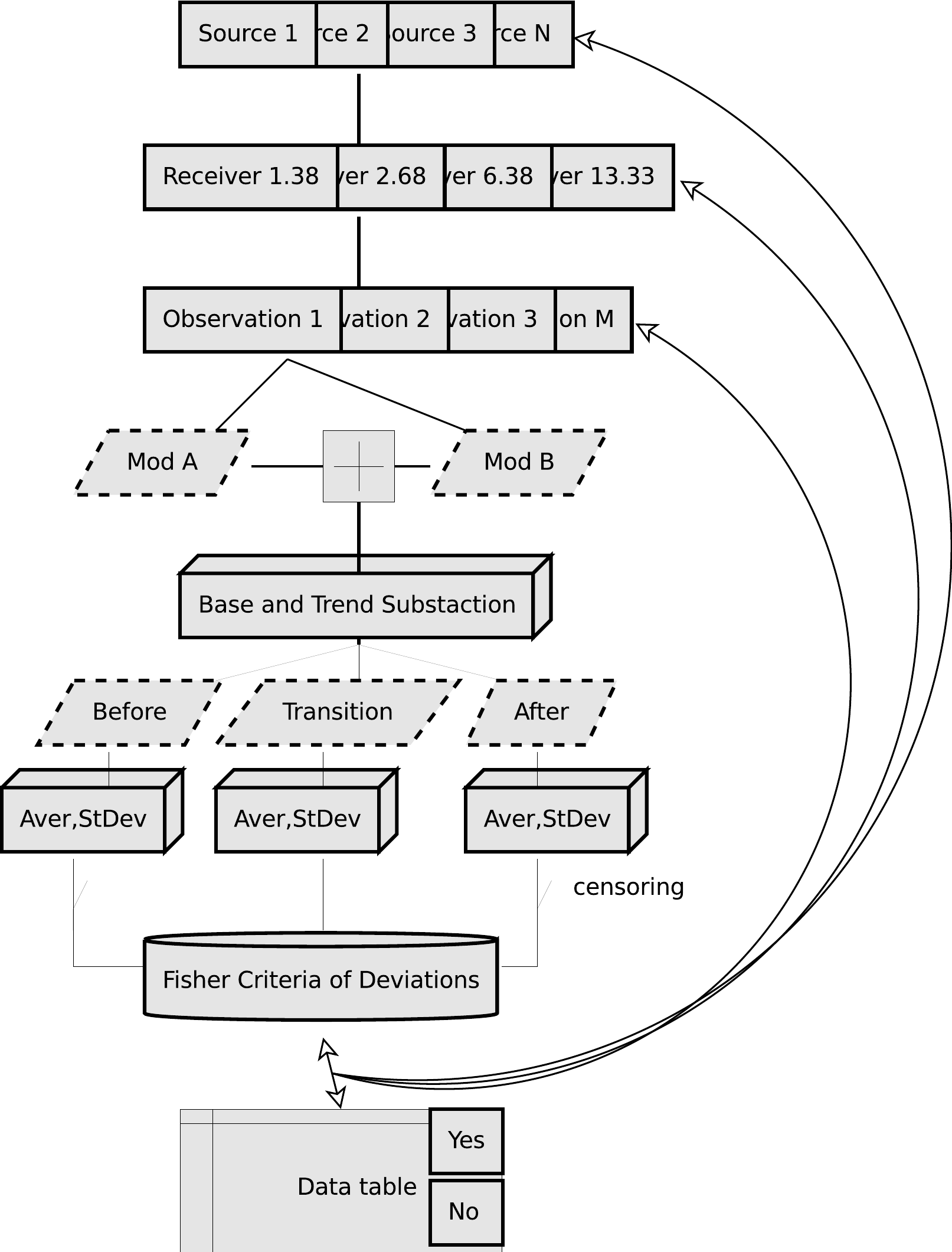}}}
\caption{Data processing pipeline.}\label{f1}
\end{figure}

We processed the observations according to the well-known method of analysis
of variance \cite{cramer1999mathematical}.
Note that the Fisher test used to compare the variance is sensitive to the
``normality'' of the original distribution of noise samples in the input signal
recording. Therefore, especially for the 13cm receiver, a rejection procedure
is mandatory - censoring of recordings affected by earthly noise.
Anomalous noises can greatly weight the right ``tail'' of the density of the sample
probability distribution, shifting its center.
So Fig.\ref{histo} shows a typical histogram of the distribution density
of the root of the sample noise variance $\sigma$ in the record J1718-3459(GJ667C)
at a wavelength of 13.33 cm.
We used well-known robust procedures to calculate the center of mass
$X_{center}$ of the sample probability distribution of the resulting
$\sigma$, based on the calculation of several rank statistics and the
selection of the median value from them (\cite{bulashev}).
A recording was considered uncorrupted by interference if:
\begin{equation}
X_{center} - G \cdot \tilde\sigma \le \sigma \le X_{center} + G \cdot \tilde\sigma
\end{equation}
Where the censoring coefficient G is calculated
by the empirical formula:
\begin{equation}
G = 1.55 + 0.8 \log(N/10)\cdot\sqrt{\epsilon - 1}
\end{equation}
Here the sample size is $N$, $\tilde\sigma$ is the estimate of
the standard deviation from $X_{center}$ , and $\epsilon$
kurtosis estimate.
The key stages of record processing are shown in the diagram in
Figure~\ref{f1}.

Choosing the analyzed source of observations
and perform the following steps in a loop for all sources. Let's transform
a multi-frequency file into a set of single-part, single-channel. Let us
choose for analysis a single-frequency record of the corresponding receiver:
1.38 cm, 2.68 cm, 6.38 cm and 13.33 cm. We choose one observation and perform
pairwise vector addition of the channel recording files from two opposite
modulation phases. We repeat this step and the next for all observations.
We cut out a part of the record of the full interval for analysis,
subtract the base level and the linear trend from it.
Thus, we discard sections of the record with the inclusion of the calibration
noise generator and their normalization by sample size, which reduces the
influence of the slow component of the temperature trend and atmospheric
noise on the resulting statistics.
We cut out three sections of records from the resulting files according to
the expressions \ref{eq:t0},\ref{eq:t1},\ref{eq:t2}.
The length and, accordingly, the number of samples for each section of
the record corresponds to the time of passage of a point source
through twice the width (at the level of 0.5 power) of the horizontal
dimension of the telescope beam pattern.
Further for each segment of the record
we calculate estimates of the mean -- $\tilde{x}$ and standard
deviation -- $\sigma$.
We enter the results in the table, including the value of the sample size.
We compare the $\sigma$ of the analyzed areas and conclude that the null
hypothesis is true: the equality of $\sigma$ according to the Fisher
criterion.
We write in the table the results of comparing $\sigma_{t1}$ with
$\sigma_{t0}$ and $\sigma_{t2}$.
We single out anomaly if the null hypothesis is rejected with a probability
of 0.99.
We repeat the calculations from the very beginning for the next observation.
We calculate $\tilde{x}$, $\sigma$ from the observation cycle for a given
radiometer wave.
We censor the resulting sample for bounces of $\sigma$ values from the
general population.
The analysis is carried out for "out-of-diagram" intervals: t0 and t2, in
order to eliminate obvious interferences that distorts the statistics.
It should be noted again that only those records affected by interference
are discarded, in which the dispersion is anomalous precisely for the control
sections before or after the source passes through the radiation pattern.
That is, these are clearly not cases of the possible
detection of radiation from the source under study.
We compute an upper limit on the $\sigma$ estimate with a probability of 0.99,
assuming that the initial $\chi^2$ is the population distribution for the
variance estimate.
We calculate the upper limit for estimating the average in the transit
interval with a probability of 0.99, assuming that the initial $\tau$ is
a Student's distribution with censoring taken into account.
We repeat the calculations starting with the choice of the receiver, and
again a full cycle for each source.
Note that the computations of each of the loops are unconditional and
therefore can be easily parallelized.
In order to convert the obtained results of the analysis from the units
of noise temperature into the spectral flux density, we processed the
calibration sources.
We tried to include observation settings for known strong radio sources
with the longest series of observations close in height.
These are: 3C48, J0237-23, 3C138, 3C147, 3C161, J1154-35, 3C286, 4C+12.50,
3C295, J1850-0101, NGC7027.
For the calculation, we used published data collected in the
CATS~\cite{cats2009} database.
Calibration method did not differ
from the standard one, based on the calculation of the effective area of
the antenna with the construction of its dependence on the height of the
installation and is described, for example, in the work of
Panov et al.~\cite{panov2019}.

\section{OBJECTS, RESULTS AND DISCUSSION  OF OBSERVATIONS}\label{disc}
Table \ref{Result} shows the results of observations of all 16 objects in our
sample. It contains the following information.

Column 1 contains coordinate name for the telescope, SIMBAD
object name of observing Star with
an exaplanets and distance to the Star in parsecs and meters -- $D$.

In column 2 there is a number of observed transits $N$.

Column 3 contains wavelength of band centre of the reseiver in centimeters.

In column 4 there is a
transit time of the source along the right ascension direction through the
antenna pattern in sidereal time seconds.

Column 5 indicates the $n$ -- number of time samples of the data in column 4.

Columns 6,7 contains the upper limit the flux density $F_s \times 10^{29}$ of the source in
mJy (1mJy $=10^{-29}$ W/(Hz$\times m^2$).
Values of the censored estimate of the upper limits of the flux density
obtained using the algorithm described in the previous section,
are calculated according
to the null hypothesis with a probability of 0.99,
calibrated against references sources with determination errors.
Because of the too high noise of the radiometers at 1.38 and 13.3 cm
wavelengths, we give in columns 8, 9, 10, 11 upper limits for the
luminosities in the most sensitive ranges.

Column 8 calculates the upper limit $P_s$ -- the power spectral density of isotropic
emitter of the stellar system. Watt value per Herz.
The following formula is used: $P_s = 4\pi D^2 F_s$.

Column 9 gives the upper limit of the star's constant luminosity $L_s$ in the
centimeter microwave range, estimated taking into account the average of
the entire observation time.
Value in units of ergs per second ($10^{-7}$ Watt).

And in column 10 its the same, but averaged over a single passage of the
source -- $L_{st}$.

Column 11 calculates the flare limit $L_{sf}$ of a Star`s luminosity in the centimeter
microwave range. Erg per second units.

The following relationships where used to estimate the values:
$L_{st} = L_{sf}/\sqrt{n}$ and $L_s = L_{st}/\sqrt{N}$;
$L_{sf} = P_s B$, where $B$ is the width of the frequency band detected by the
receiver.

The last column contains anomaly detection date for the source at the
receiver wavelength.

\clearpage
\begin{longtable*}[c]{@{}lllllllllllc@{}}
\caption{Obser\-va\-ti\-ons in April-May 2018\label{Result}}
\endfirsthead
\caption{Continuation}
\vspace{1mm}\\
\toprule[0.5pt]
\toprule[0.5pt]
\vspace{1mm}\\
1\label{col1} & 2 & 3 & 4 & 5 & \multicolumn{1}{c}{ 6 }& ~~~7 &\multicolumn{1}{c}{ 8 } & ~~~9 &
~~~10 & ~~~11 & 12\vspace{1mm}\\
\hline
\endhead
\toprule[0.5pt]
\toprule[0.5pt]
\begin{minipage}[t]{0.06\columnwidth}
Source, distance\strut
\end{minipage} &
\begin{minipage}[t]{0.06\columnwidth}
Ob\-ser\-va\-ti\-ons count\strut
\end{minipage} & \begin{minipage}[t]{0.04\columnwidth}
$\lambda$\strut
\end{minipage} & \begin{minipage}[t]{0.06\columnwidth}
Transit time\strut
\end{minipage} & \begin{minipage}[t]{0.08\columnwidth}
Discrets count\strut
\end{minipage} & \begin{minipage}[t]{0.06\columnwidth}
Upper limit of flux density\strut
\end{minipage} & \begin{minipage}[t]{0.06\columnwidth}
Flux density error\strut
\end{minipage} & \begin{minipage}[t]{0.06\columnwidth}
Upper limit of the Star's power spectral density\strut
\end{minipage}
& \multicolumn{3}{c}{\begin{minipage}[t]{0.18\columnwidth}
The upper limits of the Star's luminosity estimation at microwaves:\\
({\bf 9}) constant luminosity with whole observing average;
({\bf 10}) constant luminosity with one transit average;
({\bf 11}) flare radiation limit.\strut
\end{minipage}}
& \begin{minipage}[t]{0.11\columnwidth}
Detection date of the anomaly\strut
\end{minipage}\tabularnewline
\hline
\vspace{1mm}\\
1 & 2 & 3 & 4 & 5 & 6 & 7 &\multicolumn{1}{c}{ 8 } & ~~~9 & ~~~10 & ~~~11 & 12\vspace{1mm}\\
\hline
\vspace{1mm}\\
parsec, m & N & cm & c & n &  mJy & mJy &W/Hz& erg/s & erg/s & \multicolumn{1}{c}{ erg/s } & ~ \tabularnewline
\hline
\vspace{1mm}\\
0144-1556 & 2\vspace{1mm} & ~ & ~ & ~ & ~ & ~ & ~ & ~ & ~ & ~ & ~\tabularnewline\vspace{1mm}
Tau Cet& ~ & 1.38 & 1.54 & 26 & ~650 & 32 & ~ & ~ & ~ & ~ &
~ \tabularnewline\vspace{1mm}
3.652, 1.13E17& ~ & 2.68 & 2.96 & 51 & ~~82 &4.1 & 1.3E8 & 1.1E23&1.5E23  & 1.1E24 &
07.04.18 \tabularnewline\vspace{1mm}
~ & ~ & 6.38 & 7.12 & 122 & ~100 & 5.0 &1.6E8& 6.1E22 &8.7E22 & 9.6E23  & ~ \tabularnewline\vspace{1mm}
~ & ~ & 13.33 & 14.86 & 256 & 1500&90 & ~ & ~ & ~ & ~ & ~ \tabularnewline\hline
\vspace{1mm}\\
0428-2510 & 3\vspace{1mm} & ~ & ~ & ~ & ~ & ~ & ~ & ~ & ~ & ~ & ~ \tabularnewline\vspace{1mm}
GJ3293 & ~ & 1.38 & 1.68 & 28 & ~520&26  & ~ & ~ & ~ & ~
& ~ \tabularnewline\vspace{1mm}
20.207, 6.24E17& ~ & 2.68 & 3.26 & 56 & ~~94&4.7  &4.6E9&2.9E24 &4.9E24 & 3.7E25  &
~ \tabularnewline\vspace{1mm}
~ & ~ & 6.38 & 7.8 & 134 & ~130&6.7 & 6.4E9&1.9E24 &3.3E24  & 3.8E25  &
09.04.18 \tabularnewline\vspace{1mm}
~ & ~ & 13.33 & 16.3 & 281 & ~570&34 & ~ & ~ & ~ & ~
& ~ \tabularnewline\hline
\vspace{1mm}\\
0453-1746 & 3\vspace{1mm} & ~ & ~ & ~ & ~ & ~ & ~ & ~ & ~ & ~ \tabularnewline\vspace{1mm}
GJ180 & ~ & 1.38 & 1.56 & 26 & ~480&24 & ~& ~ & ~ & ~
& ~ \tabularnewline\vspace{1mm}
11.949, 3.69E17& ~ & 2.68 & 3.04 & 52 & ~100&5.0& 1.7E9&1.1E24 &1.9E24  & 1.4E25 &
~ \tabularnewline\vspace{1mm}
~ & ~ & 6.38 & 7.24 & 124 & ~100&5.2 & 1.7E9&5.2E23 &9.0E23  & 1.0E25
& ~ \tabularnewline\vspace{1mm}
~ & ~ &13.33 & 15.14 & 261 & ~120&74 & ~ & ~ & ~  & ~  & ~ \tabularnewline\hline
\vspace{1mm}\\
0727+0513 & 26\vspace{1mm} & ~ & ~ & ~ & ~ & ~ & ~ & ~ & ~ & ~ & ~ \tabularnewline\vspace{1mm}
GJ273 & ~ & 1.38 & 1.34 & 23 & ~420&21 & ~ & ~ & ~ & ~
& ~ \tabularnewline\vspace{1mm}
3.786, 1.17E17& ~ & 2.68 & 2.6 & 44 & ~~96&4.8 &1.7E8&3.8E22 &2.0E23 & 1.3E24  & ~
\tabularnewline\vspace{1mm}
~ & ~ & 6.38 & 6.18 & 106 & ~100&5.1 &1.7E8&1.9E22 &9.7E22 & 1.0E24
& ~ \tabularnewline\vspace{1mm}
~ & ~ &13.33 & 12.94 & 223 & ~960&48 & ~ & ~ & ~  & ~ & ~ \tabularnewline\hline
\vspace{1mm}\\
1130+0735 & 26\vspace{1mm} & ~ & ~ & ~ & ~ & ~ & ~ & ~ & ~ & ~ & ~ \tabularnewline\vspace{1mm}
K2-18 & ~ & 1.38 & 1.32 & 22 & ~390&20 & ~ & ~ & ~ & ~
& 07.04.18; 03.05.18 \tabularnewline\vspace{1mm}
38.1, 1.18E18& ~ & 2.68 & 2.58 & 44 & ~~91&4.5 &1.6E10&3.8E24 &2.0E25 & 1.3E26 &
~ \tabularnewline\vspace{1mm}
~ & ~ & 6.38 & 6.14 & 105 & ~~94&4.7 &1.6E10&1.9E24 &9.7E24 & 9.9E25
& ~ \tabularnewline\vspace{1mm}
~ & ~ & 13.33 & 12.84 & 221 & ~800&40 & ~ & ~ & ~ & ~  & ~ \tabularnewline\hline
\vspace{1mm}\\
1145+0000 & 19\vspace{1mm} & ~ & ~ & ~ & ~ & ~ & ~ & ~ & ~ & ~ & ~\tabularnewline\vspace{1mm}
K2-9 & ~ & 1.38 & 1.36 & 23 & ~420&21 & ~ & ~ & ~  & ~
& 03.05.18 \tabularnewline\vspace{1mm}
82.96, 2.56E18& ~ & 2.68 & 2.64 & 45 & ~110&5.3 &9.1E10&2.5E25 &1.0E26  & 7.2E26 &
~ \tabularnewline\vspace{1mm}
~ & ~ & 6.38 & 6.32 & 108 & ~100&5.0 &8.2E10&1.1E25 &4.7E25  & 4.9E26
& ~ \tabularnewline\vspace{1mm}
~ & ~ & 13.33 & 13.22 & 227 & ~850&43 & ~& ~ & ~ & ~ & 20.04.18 \tabularnewline\hline
\vspace{1mm}\\
1630-1239 & 27\vspace{1mm} & ~ & ~ & ~ & ~ & ~ & ~ & ~ & ~ & ~ & ~\tabularnewline\vspace{1mm}
Wolf 1061 & ~ & 1.38 & 1.48 & 25 & ~480&24 & ~& ~ & ~ & ~
& ~ \tabularnewline\vspace{1mm}
4.308, 1.33E17 & ~ & 2.68 & 2.9 & 50 & ~~97&4.9 &2.2E8&4.6E22 &2.4E23 & 1.7E24 & ~
\tabularnewline\vspace{1mm}
~ & ~ & 6.38 & 6.9 & 118 & ~~87&4.3 &1.9E8&2.1E22 &1.1E23  & 1.2E24  &
07.04.18 \tabularnewline\vspace{1mm}
~ & ~ & 13.33 & 14.44 & 248 & ~620&31 & ~ & ~ & ~ & ~
& 09.04.18 \tabularnewline\hline
\vspace{1mm}\\
1718-3459 & 26\vspace{1mm} & ~ & ~ & ~ & ~ & ~ & ~ & ~ & ~ & ~ & ~ \tabularnewline\vspace{1mm}
GJ667C  & ~ & 1.38 & 1.88 & 32 & ~530&27 & ~ & ~ & ~ & ~
& ~ \tabularnewline\vspace{1mm}
7.243, 2.23E17& ~ & 2.68 & 3.66 & 63 & ~120&6.0 &7.5E8&1.5E23 &7.6E23  & 6.0E24 &
~ \tabularnewline\vspace{1mm}
~ & ~ & 6.38 & 8.74 & 150 & ~~90&4.5 &5.6E8&5.4E22 &2.8E23 & 3.4E24
& ~ \tabularnewline\vspace{1mm}
~ & ~ & 13.33 & 18.28 & 315 & ~640&32 & ~ & ~ & ~  & ~  & ~ \tabularnewline\hline
\vspace{1mm}\\
1852+4520 & 6\vspace{1mm} & ~ & ~ & ~ & ~ & ~ & ~ & ~ & ~ & ~ & ~ \tabularnewline\vspace{1mm}
Kepler-62 & ~ & 1.38 & 1.5 & 25 & ~470&24 & ~ & ~ & ~ & ~  &
~ \tabularnewline\vspace{1mm}
301.1, 9.29E18& ~ & 2.68 & 2.92 & 50 & ~140&7.0 &1.5E12&6.9E26 &1.7E27 & 1.2E28  &
~ \tabularnewline\vspace{1mm}
~ & ~ & 6.38 & 6.96 & 120 & ~110&5.5 &1.2E12&2.7E26 &6.6E26  & 7.2E27 &
~ \tabularnewline\vspace{1mm}
~ & ~ & 13.33 & 14.54 & 250 & ~600& 30 & ~ & ~ & ~ & ~ & ~ \tabularnewline\hline
\vspace{1mm}\\
1906+4926 & 6\vspace{1mm} & ~ & ~ & ~ & ~ & ~ & ~ & ~ & ~ & ~ & ~ \tabularnewline\vspace{1mm}
Kepler-296 & ~ & 1.38 & 1.66 & 28 & ~680&34 & ~ & ~ & ~  & ~ &
~ \tabularnewline\vspace{1mm}
219.6, 6.78E18& ~ & 2.68 & 3.22 & 55 & ~140&6.9 &8.1E11&3.6E26 &8.8E26 & 6.5E27  &
~ \tabularnewline\vspace{1mm}
~ & ~ & 6.38 & 7.68 & 132 & ~120&6.1 &6.9E11&1.5E26 &3.7E26 & 4.2E27  &
~ \tabularnewline\vspace{1mm}
~ & ~ & 13.33 & 16.06 & 276 & ~620&31 & ~ & ~ & ~  & ~ & ~ \tabularnewline\hline
\vspace{1mm}\\
1916+4753 & 6\vspace{1mm} & ~ & ~ & ~ & ~ & ~ & ~ & ~ & ~ & ~ & ~ \tabularnewline\vspace{1mm}
Kepler-22 & ~ & 1.38 & 1.6 & 27 & ~580&29 & ~ & ~ & ~ & ~ &
~ \tabularnewline\vspace{1mm}
197.5, 6.1E18& ~ & 2.68 & 3.1 & 53 & ~140&7.0 &6.5E11&2.9E26 &7.1E26 & 5.2E27 &
~ \tabularnewline\vspace{1mm}
~ & ~ & 6.38 & 7.4 & 127 & ~130&6.3 &6.1E11&1.3E26 &3.9E26 & 3.6E27 &
~ \tabularnewline\vspace{1mm}
~ & ~ & 13.33 & 15.46 & 266 & ~510&25 & ~ & ~ & ~  & ~ & ~ \tabularnewline\hline
\vspace{1mm}\\
1926+3852 & 6\vspace{1mm} & ~ & ~ & ~ & ~ & ~ & ~ & ~ & ~ & ~ & ~ \tabularnewline\vspace{1mm}
Kepler-1552 & ~ & 1.38 & 1.38 & 23 & ~680&34 & ~ & ~ & ~ & ~
& 09.04.18 \tabularnewline\vspace{1mm}
769.5, 2.37E19& ~ & 2.68 & 2.68 & 46 & ~120&6.1 & 8.5E12&4.1E27 &1.0E28  & 6.8E28  &
09.04.18 \tabularnewline\vspace{1mm}
~ & ~ & 6.38 & 6.4 & 110 & ~110&6.1 &7.8E12&1.8E27 &4.5E27 &4.7E28 & ~
\tabularnewline\vspace{1mm}
~ & ~ & 13.33 & 13.36 & 230 & ~710&35 & ~ & ~ & ~  & ~ & ~ \tabularnewline\hline
\vspace{1mm}\\
1941+4228 & 26\vspace{1mm} & ~ & ~ & ~ & ~ & ~ & ~ & ~ & ~ & ~ & ~\tabularnewline\vspace{1mm}
Kepler-61 & ~ & 1.38 & 1.42 & 24 & ~490&25 & ~ & ~ & ~ & ~
& 02.05.18 \tabularnewline\vspace{1mm}
341.3, 1.05E19& ~ & 2.68 & 2.78 & 47 & ~130&6.7 & 1.8E12&4.0E26 &2.8E27 & 1.4E28 &
~ \tabularnewline\vspace{1mm}
~ & ~ & 6.38 & 6.6 & 113 & ~110&4.5 & 1.5E12 &1.7E26 &8.6E26 & 9.1E27 & ~
\tabularnewline\vspace{1mm}
~ & ~ & 13.33 & 13.84 & 238 & ~760&38 & ~ & ~ & ~ & ~ & ~ \tabularnewline\hline
\vspace{1mm}\\
1949+4659 & 27\vspace{1mm} & ~ & ~ & ~ & ~ & ~ & ~ & ~ & ~ & ~ & ~\tabularnewline\vspace{1mm}
Kepler-1229 & ~ & 1.38 & 1.56 & 26 & ~500&25 & ~ & ~ & ~ & ~  &
26.04.18 \tabularnewline\vspace{1mm}
268.2, 8.28E18& ~ & 2.68 & 3.02 & 52 & ~140&7.0 &1.2E12&2.6E26 &1.3E27  & 9.6E27 & ~
\tabularnewline\vspace{1mm}
~ & ~ & 6.38 & 7.22 & 124 & ~130&6.3 &1.1E12&1.2E26 &6.0E26 & 6.7E27
& ~ \tabularnewline\vspace{1mm}
~ & ~ & 13.33 & 15.1 & 260 & ~760&38 & ~ & ~ & ~  & ~
& ~ \tabularnewline\hline
\vspace{1mm}\\
2218-0936 & 6\vspace{1mm} & ~ & ~ & ~ & ~ & ~ & ~ & ~ & ~ & ~ & ~ \tabularnewline\vspace{1mm}
K2-72 & ~ & 1.38 & 1.46 & 25 & ~480&24 & ~ & ~ & ~ & ~ &
~ \tabularnewline\vspace{1mm}
66.51, 2.05E18& ~ & 2.68 & 2.82 & 48 & ~110&5.6 & 5.8E10&2.7E25 &6.6E25 & 4.6E26 &
~ \tabularnewline\vspace{1mm}
~ & ~ & 6.38 & 6.74 & 116 & ~~95&4.7 & 5.0E10&1.1E25 &2.8E25  & 3.0E26 &
~ \tabularnewline\vspace{1mm}
~ & ~ & 13.33 & 14.1 & 243 & ~830&42 & ~ & ~ & ~ & ~ &
~ \tabularnewline\hline
\vspace{1mm}\\
2306-0502 & 6\vspace{1mm} & ~ & ~ & ~ & ~ & ~ & ~ & ~ & ~ & ~ & ~\tabularnewline\vspace{1mm}
Trappist-1 & ~ & 1.38 & 1.4 & 24 & ~540&27 & ~ & ~ & ~ & ~ &
~ \tabularnewline\vspace{1mm}
12.47, 3.85E17& ~ & 2.68 & 2.72 & 46 & ~160&8.0 &3.0E9&1.4E24 &3.5E24 & 2.4E25 &
~ \tabularnewline\vspace{1mm}
~ & ~ & 6.38 & 6.5 & 112 & ~~94&4.7 &1.8E9&4.2E23 &1.8E24 & 1.1E25 &
~ \tabularnewline\vspace{1mm}
~ & ~ & 13.33 & 13.6 & 234 & ~570&28 & ~ & ~ & ~ & ~ &
~ \tabularnewline
\bottomrule[0.5pt]
\bottomrule[0.5pt]
\end{longtable*}

Below is a brief description of the observed stars, some results of other
observations of them in radio and other ranges, and features of the planets
orbiting  them.

{\bf Tau Cet} has a spectral class of G8V, $V = 3^m.50$, color index
$V-R = 0^m.62$, distance 3.65 pc \cite{2000A&AS..143....9W}, effective temperature $5333 \pm 99 K$,
radius $0.83 R_{\astrosun}$, luminosity $0.51 L_{\astrosun}$ \cite{2019AJ....158..138S}. The rotation
period of the star is 32d.9 - 34d.5 \cite{2011MNRAS.413L..71W, 2003A&A...397..147P}, its brightness
varied in the $V (3^m.2 - 4^m)$
and $Ic (2^m.7 - 3^m)$ bands from 1961 to 2024, while no flares of significant
amplitude were detected \cite{webobs2024, Maehara2014}. Variable radiation with flux variations from
$10^{-15}$ to $10^{-12}$ mW/$m^2$ was recorded from Tau Cet in the X-ray range of
0.2-12 keV \cite{2024yCat..36870250Q}. In the range of 0.3-8 keV the flux is
$1.24 \times 10^{-13}$ erg/($cm^2 \times$s) \cite{Wang2016}. In the radio range at
frequencies of 4.9 and 15.0 GHz the upper limits for the radiation flux were
obtained - 1.0 mJy and 11.7 $\mu$Jy respectively, and at a frequency of 34.5 GHz
the emission was registered at the level of 25.3 $\mu$Jy \cite{1989AJ.....98..279F, 2014AAS...22315118V}.
Four Earth-like planets were discovered around Tau
Cet \cite{exoplanetarchive2024}, two of them are
located on the inner and outer boundaries of the habitable zone, their
Earth similarity indices (ESI) are 0.7 and 0.77 \cite{2021AJ....161...17D}.
The star \cite{2004ApJ...609..392J} moves with $Pm\_RA = -1721.94$~mas/yr,
so the RA shift in the carriage position must be taken into account.
The height is about $30^{\circ}$ above the horizon, due to which the
capture of ground-based sources of interference is likely: thunderclouds,
aircraft. We had 2 days of observations.
Noteworthy day 04/07/2018:
An impulse enters the initial region of passage by the diagram at 2.68cm,
its Gaussian approximation is given in the table \ref{tau2}.
\begin{table}
\caption{Pulse parameters in the record Tau~Cet~e 04/07/2018, Receiver 2.68cm:
Stellar-time of center, Amplitude, Half-Width, Integral-Area.
Top line: value, bottom line - approximation error.}
\label{tau2}
\begin{tabular}{ccc}
\hline
Object: 0144-1556 &
$\lambda$: 2.68[cm] &
Date: 07/04/2018
\end{tabular}
\begin{tabular}{cccc}
\hline
[ h:m:s.s ] & [ mK ] & [ s ] & [ K*s ] \tabularnewline
\begin{minipage}[t]{0.25\columnwidth}
01:44:53.61 0.0125 \strut\end{minipage}&
\begin{minipage}[t]{0.25\columnwidth}
124.87 \\ 28.5 \strut\end{minipage}&
\begin{minipage}[t]{0.23\columnwidth}
0.091354 0.0125 \strut\end{minipage}& 0.012143 \tabularnewline
\hline
\end{tabular}
\end{table}
Its duration is slightly more than two counts and at the peak flux
density is not less than 434 mJy, which is more than $5\sigma$ outside
the passage (80.5 mJy).
It is noteworthy that there is also an outlier at 6.38cm with a spacing
along the ``stellar'' time less than a second. See table \ref{tau6}.
\begin{table}
\caption{Pulse parameters in the record Tau~Cet~e 04/07/2018, Receiver 6.38cm:
Stellar-time of center, Amplitude, Half-Width, Integral-Area.
Top line: value, bottom line - approximation error.}
\label{tau6}
\begin{tabular}{ccc}
\hline
Object: 0144-1556 &
$\lambda$: 6.38[cm] &
Date: 07/04/2018
\end{tabular}
\begin{tabular}{cccc}
\hline
[ h:m:s.s ] & [ mK ] & [ s ] & [ K*s ] \\
\begin{minipage}[t]{0.25\columnwidth}
01:44:51.81 0.0182\strut\end{minipage}&
\begin{minipage}[t]{0.25\columnwidth}
58.771 \\ 9.84 \strut\end{minipage}&
\begin{minipage}[t]{0.23\columnwidth}
0.20123 0.0182 \strut\end{minipage} & 0.012589 \\
\hline
\end{tabular}
\end{table}
Here, the pulse flux density is 229 mJy, which is more than $3\sigma$
(74 mJy). Duration in 4 counts.
It is rather strange (for a thunderstorm) that there was no interference
at 13.3cm that day, so on 04/07/2018 $\sigma$ noise in the passage:
141.451 mK, and on 04/09/2018 743.288 mK. That is, from ground interference,
an overflight of the aircraft is most likely. However, in this case,
the nature of the broadband properties of the pulse remains unclear.
There were no responses comparable with the diagram in terms of width either
on April 7th or on April 9th.
As a result, here, as for other objects (see below in Table~\ref{Result}), restrictions have
been established on the levels of flare and stationary radiation in the
most sensitive ranges of 2.7 cm and 6.4 cm for flux density and luminosity, and
for flares on a scale of 0.05 s -- 82 mJy and 100 mJy (luminosity in the
band $\sim 10^{24}$~erg/s).

{\bf GJ 3293} - spectral class M2.5, $V = 11^m.96$, color index $V-R = 0^m.33$,
distance 20.21 pc \cite{2000A&AS..143....9W}, effective temperature - 3469 K,
radius - $0.45 R_{\astrosun}$, luminosity - $0.03 L_{\astrosun}$ \cite{2019AJ....158..138S}.
The rotation
period of the star is $41^d - 50^d$ \cite{2023ApJ...954L..50E, 2017A&A...600A..13A},
its brightness from 2012 to
2024 varied in the $V (10^m.7 - 12^m.3)$ and $Ic (9^m.6 - 9^m.8)$ bands
\cite{Maehara2014}.
In observations with the Mini-MegaTORTORA system \cite{2018AN....339..375K}
from 2017 to 2022,
two outbursts with amplitudes of $3^m.2$ and $4^m.75$ (one of the maximum
for outburst stars!) were detected \cite{MMT2024}. Four planets have been
discovered
around GJ 3293, with one planet in the habitable zone with an ESI of
0.63 \cite{hwc2024}.
This star \cite{2018A&A...616A...7S} moves with $Pm\_RA = -87$~mas/yr was
observed three days April 7-9. The height above the horizon is $21^{\circ}$,
so the probability of trapping ground interferences by the antenna pattern
is also very high.
Only the 3rd day is remarkable: at 6.38cm, in the zone of passage through
the diagram directivity due to a single-sawtooth pulse an increase in
$\sigma$ noise temperature was observed.
The parameters of approximation of this pulse by two Gaussians are presented
in the table \ref{gj3293}.
\begin{table}
\caption{Double pulse parameters in the entry GJ3293d 04/09/2018, Receiver 6.38cm:
Stellar-time of center, Amplitude, Half-Width, Integral-Area.
Top line: value, bottom line - approximation error.}
\label{gj3293}
\begin{tabular}{ccc}
\hline
Object: 0428-2510 &
$\lambda$: 6.38[cm] &
Date: 09/04/2018
\end{tabular}
\begin{tabular}{cccc}
\hline
[ h:m:s.s ] & [ mK ] & [ s ] & [ K*s ] \\
\begin{minipage}[t]{0.23\columnwidth}
04:29:17.43 0.0114 04:29:18.29 0.009\strut\end{minipage}&
\begin{minipage}[t]{0.22\columnwidth}
97.36 \\1.42\\122.91\\4.20 \strut\end{minipage} &
\begin{minipage}[t]{0.22\columnwidth}
1.299 \\0.0114\\0.4389\\0.009 \strut\end{minipage} &
\begin{minipage}[t]{0.22\columnwidth}
0.1347 \\~\\ 0.0574 \strut\end{minipage}\\
\hline
\end{tabular}
\end{table}
If the radio emission pulse is of a non-interference nature, then the
density value flux per pulse is slightly above 2$\sigma$ of background noise.

{\bf GJ 180} - spectral class M3V, $V = 10^m.90$, color index $V-R = 1^m.02$,
distance 11.95 pc \cite{2000A&AS..143....9W}, effective temperature - 3556 K, radius -
$0.41 R_{\astrosun}$,
luminosity - $0.03 L_{\astrosun}$ \cite{2019AJ....158..138S}. The rotation period of the star
is determined to be within $53^d - 71^d$ \cite{2024A&A...684A...9S}, its brightness from
2005 to 2024 varied in the V band from $10^m.9$ to $12^m.5$ \cite{Maehara2014,2009ApJ...696..870D},
and flares with an amplitude of no more than 1m were recorded \cite{Maehara2014}.
GJ 180, according to observations in the range of 0.2-2.3 keV, has a
luminosity of $7.45 \times 10^{19}$~W, its radio emission was
not detected. GJ 180 has three exoplanets \cite{exoplanetarchive2024},
two of which are located
in the habitable zone, with ESIs of 0.70 and 0.48 \cite{hwc2024}.
A star system GJ 180 \cite{2020ApJS..246...11F} have a
$Pm\_RA = 408$~mas/yr \cite{2018A&A...616A...1G}.
Three passages were observed, and no anomalous increases in $\sigma$ were
recorded.

{\bf GJ 273} is a star of spectral class M3.5V,
$V = 9^m.87$, color index $V-R = 1^m.17$, distance 3.79 pc
\cite{2000A&AS..143....9W},
effective temperature 3382 K, radius $0.32 R_{\astrosun}$,
luminosity $0.01 L_{\astrosun}$
\cite{2019AJ....158..138S, 2017A&A...602A..88A}.
The rotation period of the star is about 100 days
\cite{2017A&A...602A..88A}, its brightness
from 2010 to 2024 varied in the V band from $9^m.6$ to $10^m.5$, while no
outbursts were registered \cite{Maehara2014}. In the range of 0.2-2 keV,
the star
has a luminosity of $6.62 \times 10^{19}$~W
\cite{2022A&A...661A..23F}. Its radio emission at
frequencies from 74 MHz to 4.9 GHz was absent \cite{1994AJ....107.1829C}.
Two exoplanets have
been discovered around GJ 273, one of which, with an ESI of 0.85, is
located in the habitable zone \cite{hwc2024}.
Leuthen's Star \cite{1943ApJ....97..381V}
moves with $Pm\_RA = 572$~mas/yr \cite{2018A&A...616A...1G}.
We observed 26 passages without any significant increase in $\sigma$ noise.

\begin{figure}
\centerline{\vbox{\includegraphics[width=8.0cm]{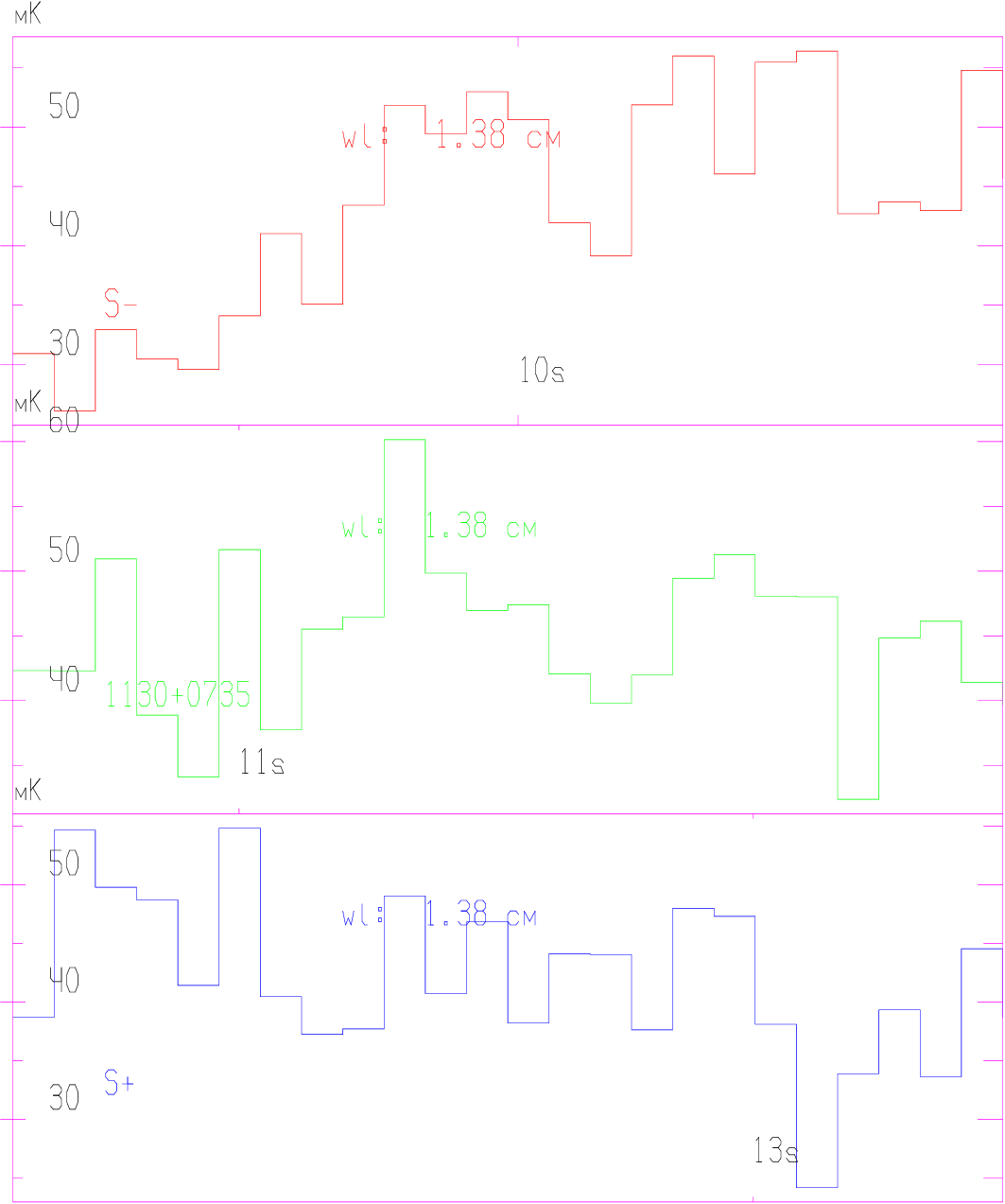}}}
\caption{Sample $\sigma$ of 26 records the K2-18 Star for 1.38cm receiver.
The graphs from top to bottom show the changes in sample $\sigma$ before,
during, and after passing through the antenna diagram.}
\label{fk218-138}
\end{figure}

\begin{figure}
\centerline{\vbox{\includegraphics[width=8.0cm]{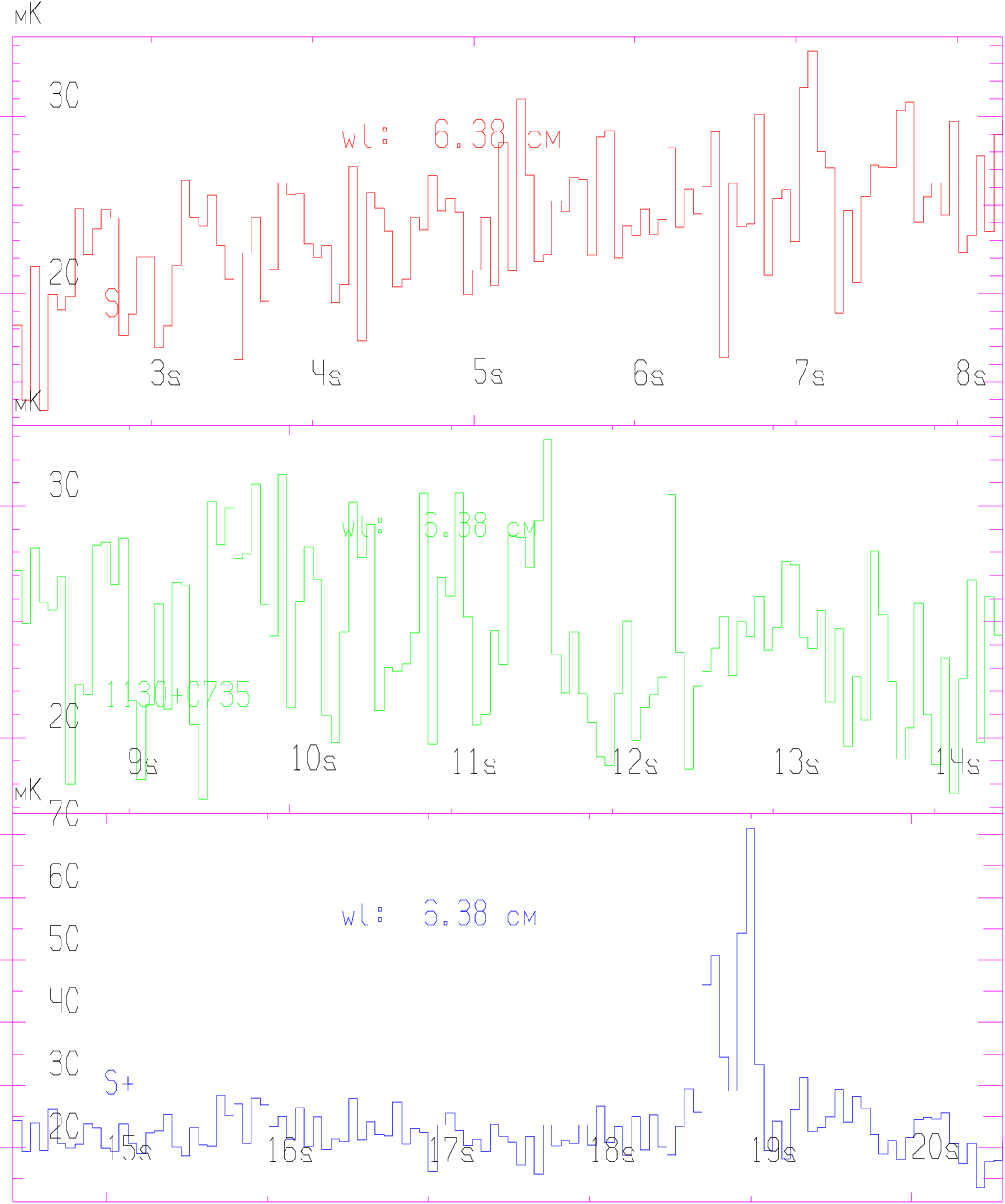}}}
\caption{Sample $\sigma$ of 26 records the K2-18 Star for 6.38cm receiver.
The graphs from top to bottom show the changes in sample $\sigma$ before,
during, and after passing through the antenna diagram.}
\label{fk218-638}
\end{figure}

{\bf K2-18} is a star of spectral class
dM2.5, $V = 13^m.50$, color index $V-R = 0^m.26$, distance 38.1 pc
\cite{2000A&AS..143....9W},
$Pm\_RA = -87.377~$mas/yr~\cite{2018A&A...616A...1G},
effective temperature - 3457K, radius - $0.45 R_{\astrosun}$,
luminosity - $0.03 L_{\astrosun}$ \cite{2019ApJ...887L..14B}.
The rotation period of the star is $39^d.63$ \cite{2018AJ....155..257S},
its brightness from 2005 to 2024 varied in the $V (11^m.3 - 12^m.4)$ and
$Ic (12^m.95 - 13^m.87)$ bands, and flares with an amplitude of up
to $1^m.6$ were recorded \cite{Maehara2014, 2009ApJ...696..870D}.
There are no data on X-ray emission, radio emission was not registered
\cite{2011MNRAS.413L..71W, 1994AJ....107.1829C, becker1995first}.
Two exoplanets \cite{exoplanetarchive2024} have been discovered around
K2-18, one of which with an ESI of 0.70 is located in the habitable
zone \cite{hwc2024}.
The sources 2MASS J11301676+0733404 Mag 15.5(g) and the quasar QSO B1127+078
(down $\delta$ by $2'$), 113017.3818+073213.0294 (J2000) Mag 14.6 may fall
into the RATAN-600 diagram lower in $\delta$ (K).
At a wave of 6.38cm, the vertical half-width of the diagram in this case is
about $2^{\circ}$.
What could affect the determination of the upper limit of detection of the
density of the radio emission flux from the star itself.
Processing of 26 records showed that without compression (dt=0.054c),
almost all days of $\sigma$ before, during and after the passage of the
radiation pattern are identical according to the Fisher criterion with a
probability of more than 95\%.
The exception is the recording of April 20, 2018, where, in the last interval
of the analysis, an interference spike was detected at a wavelength of 6.38
cm with a duration of 0.375 s, much shorter than the half-width of the
diagram, and an intensity of $4\sigma$ of background noise.
On a wave of 1.38cm, anomalous days: 04/07/2018 and 05/03/2018, the sigma
in the passage increased 1.3 and 1.8 times, respectively.
Confusion interference hypothesis is confirmed by a
comparison of the changes in the sample $\sigma$ when passing through the
antenna diagram for receivers of 1.38 cm (Fig.\ref{fk218-138}) and 6.38 cm
(Fig.\ref{fk218-638}), respectively, calculated from 26 observations.
The interference is significantly greater in the 6.38 cm receiver channel
and despite the proportionally larger size of the horizontal section of the
diagram, the duration of the interference pulse remains significantly shorter.
This is also typical for an orbiting satellite crossing the antenna pattern.


{\bf K2-9} has a spectral class of M2.5V, $V = 15^m.84$, color index $V-R = 0^m.51$,
distance 82.96 pc \cite{2000A&AS..143....9W}, effective temperature 3528 K,
radius $0.32 R_{\astrosun}$, luminosity $0.02 L_{\astrosun}$
\cite{2019AJ....158..138S}. The rotation period of the star is in the
range of $17^d.55 - 23^d.9$ \cite{2023ApJ...954L..50E}, its brightness
from 2005 to 2013 in the V band was $14^m.84 \pm 0^m.03$
\cite{2009ApJ...696..870D}, while flares of significant amplitude were
not registered. Radio emission was not detected \cite{2011MNRAS.413L..71W},
there are no data on X-ray emission. K2-9 has one exoplanet
\cite{exoplanetarchive2024} with an ESI of 0.70 located in the habitable
zone \cite{hwc2024}.
Source in 2MASS catalog
J11450348+0000190~\cite{2003IAUS..211..189K} moves with $Pm\_RA = -170.752$~mas/yr.
19 passages were observed.
During the passage on 04/20/2018 in the 13.3 cm band, an interference in the
form of a double pulse was registered, the approximation parameters of which
are given in the table \ref{K2-9}.
\begin{table}
\caption{Double pulse parameters in record K2-9b 04/20/2018, Receiver 13cm:
Stellar-time of center, Amplitude, Half-Width, Integral-Area.
Top line: value, bottom line - approximation error.}
\label{K2-9}
\begin{tabular}{ccc}
\hline
Object: 1145-0000 &
$\lambda$: 13.33[cm] &
Date: 20/04/2018
\end{tabular}
\begin{tabular}{cccc}
\hline
[ h:m:s.s ] & [ mK ] & [ s ] & [ K*s ] \\
\begin{minipage}[t]{0.23\columnwidth}
11:46:04.08 0.0021 11:46:04.94 0.0005\strut\end{minipage}&
\begin{minipage}[t]{0.25\columnwidth}
3522 \\43.4\\14183\\23.3 \strut\end{minipage} &
\begin{minipage}[t]{0.22\columnwidth}
0.283 \\0.002\\0.52759\\0.0005 \strut\end{minipage} &
\begin{minipage}[t]{0.22\columnwidth}
1.0616 \\~\\ 7.965 \strut\end{minipage}\\
\hline
\end{tabular}
\end{table}
Since the shape of the noise fits well into the highly compressed in time
section of the radiation pattern with offset from the focus, then most likely
this is the response to the passage of an aircraft in the near zone of the
telescope.
In the case of a non-interference nature, the flux density of the 1st pulse
is $\sim 0.55$~Jy, and the 2nd pulse is $\sim 51$~Jy.
At a wavelength of 1.38cm during the passage of $\sigma$ on May 3, 2018,
the noise increased by a factor of 1.44.

{\bf Wolf 1061} is a single star of the BY Dra type. Spectral class
M3V, $V = 10^m.07$, color index $V-R = 1^m.16$, distance 4.31 pc \cite{2000A&AS..143....9W},
$Pm_RA = -94.212$ mas/yr \cite{2018A&A...616A...1G}, effective temperature
3309 K, radius $0.32 R_{\astrosun}$, luminosity $0.01 L_{\astrosun}$
\cite{2019AJ....158..138S}. The rotation periods of the star are $95^d$ and
$119^d$ \cite{2017A&A...602A..88A, 2019A&A...621A.126D}, its brightness
from 2011 to 2024 in the V, B and Ic bands had values of $10^m.14 \pm 0^m.11$,
$11^m.59 \pm 0^m.26$ and $7^m.48 \pm 0^m.09$, while no outbursts were
registered \cite{2009ApJ...696..870D}. In the range of 0.2-2 keV, the
luminosity of the star was about $3.52 \times 10^{19}$ W \cite{2022A&A...661A..23F}.
According to \cite{Ortiz2024} in the range of 4 - 8 GHz the luminosity was
less than $6.7 \times 10^{11}$ erg/(s$\times$Hz). Three exoplanets
\cite{exoplanetarchive2024} were discovered around Wolf~1061, one of
which with ESI 0.80 is located in the habitable zone \cite{hwc2024}.
RATAN-600 observed 27 passages.
At a wavelength of 13cm on April 9, 2018 and April 26, 2018, $\sigma$ of
noise increased by factors of 1.7 and 3, respectively, in the latter case,
most likely not associated with the source.
At a wave of 6.38cm on April 7, 2018, an anomalous increase to 1.84$\sigma$
in transit.
This is higher than the Fisher constraint for the null hypothesis
(probability 0.99).

{\bf GJ 667 C} - spectral class M1.5V, $V = 10^m.22$, color index $V-R = 0^m.17$,
distance 7.24 pc \cite{2000A&AS..143....9W}, $Pm_RA = 1129.76$~mas/yr
\cite{2018A&A...616A...1G}, effective temperature - 4819~K~\cite{2019AJ....158..138S},
radius - $0.52 R_{\astrosun}$, luminosity - $0.13 L_{\astrosun}$ \cite{2015arXiv151001731T}.
The rotation period of the star is $103^d.9$ \cite{2015MNRAS.452.2745S}.
In the range of 0.2-2 keV, the luminosity of the star is $6.48 \times 10^{27} 10^{-7} W$
\cite{2022A&A...661A..23F}. The upper limit of the radio luminosity at
frequencies of 4 - 8 GHz is $2.1 \times 10^{12}$ erg/(s$\times$Hz)
\cite{Ortiz2024}. Five exoplanets have been discovered around GJ 667 C
\cite{exoplanetarchive2024}, three with ESI0.80, 0.60, 0.76 and located
in the habitable zone \cite{hwc2024}.
26 passages were observed at RATAN-600.
Height is only $11^\circ$ above the horizon.
Strong interference at 13.3cm on April 11,12,18,22 and 30th.
To obtain homogeneous samples for the 13.3cm radiometer, censoring was
applied. No other anomalous events were observed.

{\bf Kepler-62} has a spectral class of G5, $V = 13^m.80$, color
index $V-R = 0^m.18$, distance 301 pc \cite{2000A&AS..143....9W},
$Pm\_RA = -25.15$~mas/yr \cite{2018A&A...616A...1G}, effective temperature
4842 K, radius $0.73 R_{\astrosun}$, luminosity $0.26 L_{\astrosun}$
\cite{2019AJ....158..138S}. The rotation periods of the star are $29^d.90$
and $35^d.99$ \cite{2018MNRAS.474.2094A, 2015ApJ...801....3M},
its brightness varied in the V, R, I bands from $13^m.7$ to $14^m.4$
from 2004 to 2024. Over 150 outbursts were recorded, the amplitude of
which did not exceed $0^m.5$ \cite{2009ApJ...696..870D,2006PASP..118.1407P,2021AJ....162...98B,
2011MNRAS.418...96B,Masci2019,2023arXiv230403791H}. Five exoplanets
\cite{exoplanetarchive2024} have been discovered around Kepler-62,
two of which with ESI 0.83 and 0.68 are located in the habitable
zone \cite{hwc2024}.
On RATAN-600, no abnormal increase in $\sigma$ noise was detected after 6
passes.

{\bf Kepler-296} has a spectral class of M2V, $V = 16^m.45$, color index
$V-R = 0^m.64$, distance 220 pc \cite{2000A&AS..143....9W},
$Pm\_RA = 2~$mas/yr \cite{2018A&A...616A...1G}, effective temperature 3544~K
\cite{2018A&A...616A...1G}, radius $0.37 R_{\astrosun}$ \cite{2016ApJ...822...86M},
luminosity $0.03 L_{\astrosun}$ \cite{2015ApJ...800...99T}.
The rotation periods of the star are determined to be $35^d.86$ and
$39^d.36$ \cite{2013ApJ...775L..11M,2013MNRAS.436.1883W}. The change
in the brightness of the star from 2009 to 2024 in the bands close
to V, R, I did not exceed $0^m.2$ \cite{Masci2019},
25 low-amplitude flares were recorded \cite{2021AJ....162...98B}.
Five exoplanets have been discovered around Kepler-296 \cite{exoplanetarchive2024},
two in the habitable zone have ESI of 0.85 and 0.66 \cite{hwc2024}.
We observed 6 passages. No anomalous increases in $\sigma$ noise were found
at RATAN-600.

{\bf Kepler-22} has a spectral class of G5, $V = 11^m.81$, color index
$V-R = 0^m.17$, distance 198 pc \cite{2000A&AS..143....9W},
$Pm\_RA = -39.7$~mas/yr \cite{2018A&A...616A...1G}, effective temperature
$5596 \pm 61 K$, radius $0.87 R_{\astrosun}$ \cite{2023A&A...677A..33B},
luminosity $0.65 L_{\astrosun}$ \cite{2019AJ....158..138S}. The rotation
periods of the star are determined to be $19^d.25$ and $25^d.22$
\cite{2021ApJS..255...17S,2018MNRAS.474.2094A}. The change in the brightness
of the star from 2005 to 2024 in the bands close to V and R did not
exceed $0^m.2$ \cite{2011MNRAS.418...96B, 2023arXiv230403791H}, but 64
low-amplitude outbursts were recorded \cite{2021AJ....162...98B}.
Kepler-22 is orbited by one planet in the habitable zone with ESI
0.72 \cite{hwc2024}.
We observed 6 passages. No anomalous increases in $\sigma$ noise were found
at RATAN-600.

{\bf Kepler-1552} has a spectral class of K2.5V, $V = 14^m.72$, color index
$V-R = 0^m.10$, distance 770 pc \cite{2000A&AS..143....9W},
$Pm\_RA = 7$~mas/yr \cite{2020yCat.1350....0G}, effective temperature is
$4871 \pm 107$ K, radius is $1.11 R_{\astrosun}$, luminosity is
$0.63 L_{\astrosun}$ \cite{2019AJ....158..138S}. The rotation period of
the star is $16^d.94$ \cite{2018MNRAS.474.2094A}. The brightness
variation of the star from 2009 to 2024 in the bands close to B, V and R
did not exceed $0^m.2$ \cite{2006PASP..118.1407P,2021AJ....162...98B,
Masci2019,2023arXiv230403791H,2022yCat.1355....0G}. No flares of
significant amplitude were detected, with the exception of one event
with an amplitude of $3^m.48$ in the RP band according to Gaia data
\cite{2022yCat.1355....0G}. One exoplanet with an ESI of 0.7 was
discovered around Kepler-1552, which is located in the habitable zone
\cite{hwc2024}.
The star were observed in 6 passages at RATAN-600.
An anomaly was registered in the 1.38cm band on 04/09/18, $\sigma$ of
noise in the passage increased by 2 times. In the 2.68cm band, the
increase was 1.3 times.

{\bf Kepler-61} - spectral class M0V, $V = 15^m.17$, color index $V-R = 0^m.11$,
distance 341 pc \cite{2000A&AS..143....9W}, $Pm\_RA = -2.255$~mas/yr
\cite{2020yCat.1350....0G}, effective temperature - 4148~K, radius -
$0.70 R_{\astrosun}$, luminosity - $0.13 L_{\astrosun}$ \cite{2019AJ....158..138S}.
The rotation periods of the star are $17^d.12$ \cite{2013A&A...560A...4R} and $59^d.88$
\cite{2016AJ....151...68K}, brightness variations from 2004 to 2024 in
bands close to V and R did not exceed $0^m.05$
\cite{2006PASP..118.1407P,2021AJ....162...98B,Masci2019,2023arXiv230403791H}.
According to Kepler data \cite{2021AJ....162...98B}, about 100 low-amplitude
flares were detected. There is a planet with ESI 0.72
\cite{exoplanetarchive2024,2013AAS...22140707B} at the inner edge of
the habitable zone.
26 passages were observed at RATAN-600.
Anomalous increase in $\sigma$ of noise in the 1.38cm band  by 1.5 times was
detected 05/02/18.

{\bf Kepler-1229} - spectral type M1V, $V = 16^m.23$, color index $V-R = 0^m.40$,
distance 268 pc \cite{2000A&AS..143....9W}, $Pm\_RA = 21.6$~mas/yr
\cite{2020yCat.1350....0G}, effective temperature 3774~K, radius
$0.57 R_{\astrosun}$, luminosity $0.06 L_{\astrosun}$ \cite{2019AJ....158..138S},
rotation period $17^d.63$ \cite{2017AJ....154..264T}. The change in the
brightness of the star from 2006 to 2024 in the bands close to B, V and R
did not exceed $0^m.07$ \cite{2006PASP..118.1407P,2021AJ....162...98B,
Masci2019,2023arXiv230403791H,2022yCat.1355....0G}. About 70 low-amplitude
flares were recorded in Kepler observations \cite{2021AJ....162...98B}, and
two powerful ones - $<3^m$ and $>4^m$ in Gaia observations
\cite{2022yCat.1355....0G}. Kepler-1229's radio emission, based on
independent processing of observations from surveys
\cite{2019yCat.8102....0H,2017A&A...598A..78I,1998AJ....115.1693C,
1998AJ....115.1693C,1994AJ....107.1829C,becker1995first}, does not exceed
statistical background fluctuations. The star has one planet
\cite{exoplanetarchive2024} in the habitable zone with an ESI of 0.62
\cite{hwc2024}.
27 passages were observed at RATAN-600.
Anomalous increase in $\sigma$ of noise in the 1.38cm band by 1.6 times was
detected 04/26/18.

{\bf K2-72} has a spectral class of M2V, $V = 15^m.04$, color index $V-R = 0^m.78$,
distance 67 pc \cite{2000A&AS..143....9W}, $Pm\_RA = 195.855$~mas/yr,
effective temperature $3498 \pm 157 K$, radius $0.33 R_{\astrosun}$,
luminosity $0.02 L_{\astrosun}$ \cite{2019AJ....158..138S}, rotation period
$38^d.47$. The brightness variations of the star from 2005 to 2024 in the
bands close to V and R did not exceed $0^m.05$ \cite{2009ApJ...696..870D,
2021AJ....162...98B,2023arXiv230403791H}, six low-amplitude flares were
recorded and one with an amplitude of $0^m.9$ in the V band. K2-72 has
four exoplanets, one of which with an ESI of 0.87 is located near the inner
boundary of the habitable zone \cite{exoplanetarchive2024,hwc2024}.
6 passages were observed at RATAN-600.
No anomalous increases in $\sigma$ of noise were recorded.

{\bf TRAPPIST-1} is a low-mass star, with spectral class M7.5e, $V = 18m.80$,
color index $V-R = 2^m.33$, distance 13 pc \cite{2000A&AS..143....9W},
$Pm\_RA = 890$~mas/yr~\cite{2003yCat.2246....0C}, effective temperature
- 2566~K, radius - $0.12 R_{\astrosun}$, luminosity - $0.0006 L_{\astrosun}$
\cite{2021PSJ.....2....1A}, rotation period $1^d.40$ \cite{2016Natur.533..221G}.
According to \cite{2009ApJ...696..870D}, the brightness variations of
the star from 2005 to 2013 in the V band were $\pm 0^m.350$, a flare
at $4^m.6$ was detected. From 2016 to 2017 in the Kp band the brightness
was at the level of $12^m.917 \pm 0^m.355$ \cite{2016ApJS..224....2H}, a
flare with an amplitude of $3^m.7$ was recorded. According to \cite{Ortiz2024}
in the range of 4 - 8 GHz the upper limit for the luminosity was
$6.7 \times 10^{11}$ erg/(s$\times$Hz). The X-ray emission of the star in
the ranges of $0.3-10 keV$ is $1.04 \times 10^{-14}$ \cite{2020ApJS..247...54E}
and 0.2-12 keV is $2.00 \times 10^{-14}$ \cite{2024yCat..36870250Q} mW/$m^2$.
Seven planets \cite{exoplanetarchive2024} were discovered around TRAPPIST-1,
four of them with ESI of 0.91, 0.85, 0.68 and 0.58, located in the habitable
zone \cite{hwc2024}.
6 passages were observed at RATAN-600.
No anomalous increases in $\sigma$ of noise were
found.
\section{CONCLUSION}\label{concl}
In April-May 2018, the RATAN-600 radio telescope carried out monitoring
observations of 16 solar-type stars with planetary orbits in habitable
zones with Earth similarity indices in the range of 0.6-0.9.
For recording of radio emission, radiometers of four
ranges were used: 1.38, 2.7, 6.4, 13.3 cm with an increased sampling
frequency of $\sim 18$~Hz. The observational data were processed and
the monitoring results were analyzed.
An algorithm for processing information obtained in the mode of the
source passing through the antenna pattern was used, based on dispersion
analysis with censoring of interference. In this case, a correction of the
passage time for the secular shift of the right ascension of the sources
due to their relative proximity was taken into account.
Eleven unidentified events of anomalous excess of the standard deviation
of noise from stationary values were detected. For the most sensitive
receivers of the 2.7 and 6.4 cm ranges, their amplitude was $\sim 100$~mJy
with 99\% probability.
The nature of these anomalies indicates their noise and/or terrestrial
(thunderstorms, aircraft, satellites) origin.
After excluding anomalous events, the upper limits of the flux density
for flares on a scale of 0.05 s were 80 - 150 mJy  (microwave luminocity --
$10^{24} - 10^{30}$ erg/s, and in a steady state with a luminosity
of $10^{22} - 10^{29}$ erg/s).
However, radio emission from the observed stars was not detected.
Apparently, the absence of activity in the radio range indicates a low
level of energy release associated with it in the coronas of these stars,
which could pose a danger to exoplanetary biota.
Nevertheless, there remains the possibility of the realization of
non-stationary phenomena in sun-like stars, manifested in the generation
of high-frequency radiation and cosmic rays, and preventing the emergence
and preservation of life on their planetary systems.
To study such effects, research is needed in other
ranges, including optical.
It should be noted that the limitations for the luminosity of radio
emissions of different durations may also apply to signals from
extraterrestrial civilizations (EC) living on planets near the studied stars.
In this case, it can be argued that the power of hypothetical EC transmitters
does not exceed $2\times 10^9 - 10^{15}$ W.
Nevertheless, there remains the possibility of the realization of
non-stationary phenomena in sun-like stars, manifested in the generation
of high-frequency radiation and cosmic rays, and preventing the emergence
and preservation of life on their planetary systems.

\section*{ACKNOWLEDGMENTS}
The work was carried out within the framework of the state assigment of SAO of
RAS in the part "Conducting fundamental scientific research".
We are grateful to the Directorate of the SAO RAS for allocating the
observation time of the RATAN-600 telescope from the reserve.
\section*{CONFLICT OF INTEREST}
The authors declare no conflict of interest.
\bibliographystyle{AstroBull}
\bibliography{References}
\end{document}